\documentstyle[12pt]{article}
\begin{document}

\renewcommand{\thefootnote}{\fnsymbol{footnote} }

\begin{center}
{\bf Exponential Decay of Wavelength in a Dissipative System }\\\vspace{1cm}

Li Hua Yu\\

National Synchrotron Light Source, Brookhaven National Laboratory, N.Y.11973 %
\vspace{0.4cm}

Abstract\\
\end{center}

Applying a technique developed in a recent work[1] to calculate wavefunction
evolution in a dissipative system with Ohmic friction, we show that the
wavelength of the wavefunction decays exponentially, while the Brownian
motion width gradually increases. In an interference experiment, when these
two parameters become equal, the Brownian motion erases the fringes, the
system thus approaches classical limit. We show that the wavelength decay is
an observable phenomenon.

\newpage

1. Introduction. The simplest example of a dissipative system, an harmonic
oscillator coupled to an environment of a bath of harmonic oscillators, has
been the subject of extensive studies [2-4] (see [1] for more references).
In a recent paper[1], we obtained a simple and exact solution for the wave
function of the system plus the bath, in the Ohmic case. It is described by
the direct product in two independent Hilbert spaces. One of them is
described by an effective Hamiltonian, the other represents the effect of
the bath, i.e., the Brownian motion, thus {\it clarifying the meaning of the
wave function of the effective Hamiltonian after the bath variables are
eliminated}, establishing a connection between two different approaches to
the dissipative system problems: the effective Hamiltonian approach, and the
system plus bath approach.

In this paper we shall apply this result to a specific case to show how the
wavelength of the wavefunction decays exponentially while the Brownian
motion width increases at the same time. We shall show that in an experiment
with two overlapping wave packets interfering with each other, the
interference term of the probability density decreases rapidly {\it when the
wavelength of the wavefunction of the effective Hamiltonian reduces to be
equal to the Brownian motion width.} Therefore, our result provides a
transparent picture about how a quantum system, when interacts with
environment and hence becomes a dissipative system, gradually evolves into a
classical system.

First, we derive a simple formula for the probability density. Next, we
apply our method to the case of an initial gaussian wave packet displaced
from the center of the parabolic potential well, and show that the
wavelength of the effective Hamiltonian decreases exponentially while the
Brownian motion width increases. Then, we use the results to calculate the
interference term, assuming the initial state is a superposition of two wave
packets with minimum uncertainty, one of them is centered at the origin, the
other is displaced from the center. This problem has been studied by
A.O.Caldeira and A.J.Leggett (CL)[4]. However, the result does not show the
expected fringe wavelength decay. This puzzle stimulated our further
analysis given in the last section, which {\it explains why the wavelength
decay is not observable in this well studied case, and shows that if the
initial wave packet width is different from that with the minimum
uncertainty, the wavelength decay becomes an observable phenomenon} . We
shall use the same notation as the previous paper, and quote formulas from
there.

2. The probability density. We consider an harmonic oscillator of frequency $%
\omega _0$, linearly coupled to a bath of harmonic oscillators of frequency $%
\omega _j$ with Ohmic spectral density distribution. The Hamiltonian of this
system is given in reference [1]:

\begin{equation}
H=\frac{p^2}{2M}+\frac 12M(\omega _0^2+\Delta \omega
^2)q^2+q\sum_jc_jx_j+\sum_j\left( \frac{p_j^2}{2m_j}+\frac 12m_j\omega
_j^2x_j^2\right) \;\;\;.
\end{equation}
It is well known that the coupling to the bath introduces damping to the
harmonic oscillator [1]. The damping rate is $\eta $ and the system
frequency is shifted to $\omega =(\omega _0^2-\eta ^2/4)^{1/2}$ by the
damping. In the previous paper [1] we showed that the wavefunction of the
system plus the bath, in the Schoedinger representation, at time t, can be
written as: 
\begin{equation}
\Psi (q,\{\xi _j\},t)=\psi (q-\sum_j\xi _j,t)\,\prod_{j=1}^N\chi _j(\xi
_j,t)\,\,,
\end{equation}
where the wavefunction $\psi (Q,t)$ is a solution of the effective
Hamiltonian eq.(23) of [1]:

\begin{equation}
H_Q=e^{-\eta t}\frac{P^2}{2M}+\frac 12M\omega _0^2e^{\eta t}Q^2\,\,\,,
\end{equation}
while $\xi _j=b_{j1}(t)x_{j0}+b_{j2}(t)\dot x_{j0}\,\,\,$is the contribution
of the j'th bath oscillator to the Brownian motion with $x_{j0},\dot x_{j0}$
its initial position and velocity. The coefficients $b_{j1}(t)$ and $%
b_{j2}(t)$ are well known in elementary physics[1]. The function $\chi
_j(\xi _j,t)\,$ is given by $\chi _j(\xi _j,t)=<\theta _{\xi _j}|\chi _{j0}>$%
, where $\chi _{j0}(x_{j0})$ is the initial state of the bath oscillator,
and $|\theta _{\xi _j}>$ is an eigenstate of the operator $\xi _j$ with
eigenvalue $\xi _j$, determined to within an arbitrary phase, which depends
on only $\xi _j$:

\begin{equation}  \label{eqthetaj}
\theta _{\xi _j}(x_{j0},t)=\left( \frac{m_j}{2\pi \hbar b_{j2}}\right)
^{\frac 12}exp\left[ -\frac{im_j}{2\hbar b_{j2}}\left(
b_{j1}x_{j0}^2-2x_{j0}\xi _j\right) \right] .
\end{equation}

We are interested in the probability density time evolution if initially the
bath is in an equilibrium at temperature T. Assuming initially the j'th bath
oscillator is in an excited state $n_j$, we write the initial wavefunction
of the this oscillator as $\chi _j\left( x_{j0}\right) =\chi _j^{\left(
n_j\right) }\left( x_{j0}\right) $. Then, the probability density for
finding the system at position $q$ and finding the contribution of the bath
oscillator j to the Brownian motion to be $\xi _j$, is:

\begin{equation}  \label{desity_ksi}
\rho (q,\{\xi _j\};t)=|\psi (q-\sum_j\xi _j,t)|^2\prod_{j=1}^N \rho _j(\xi
_j;t)\,,
\end{equation}
with the distribution of $\xi _j$ at time t:

\begin{center}
\begin{equation}
\rho _j(\xi _j;t)\,\equiv \left( \sum_{n_j=0}^\infty |\chi _j^{\left(
n_j\right) }(\xi _j,t)\,|^2e^{-\beta (n_j+\frac 12)\omega _j\hbar }\right)
/Z_j\;,  \label{eqrj}
\end{equation}
\end{center}

where $\beta =1/kT$, and the partition function is

\begin{equation}
Z_j=\ \sum_{n_j=0}^\infty e^{-\beta (n_j+\frac 12)\omega _j\hbar }=\frac
1{2\sinh (\frac{\beta \hbar \omega _j}2)}.
\end{equation}
Using $\chi _j(\xi _j,t)=<\theta _{\xi _j}|\chi _{j0}>$, we find $\rho
_j(\xi _j;t)\,=<\theta _{\xi _j}|\rho _{j0}|\,\theta _{\xi _j}>$, where $%
\rho _{j0}$ is the initial density matrix of the j'th bath oscillator. The
matrix elements of $\rho _{j0}$ is well known in quantum statistics text
book[3]:

\begin{center}
\begin{equation}  \label{eqrj0}
\rho _{j0}(x_{j0},x_{j0}^{\prime })=\left( \frac{m_j\omega _j\tanh (\frac{%
\beta \hbar \omega _j}2)}{\hbar \pi }\right) ^{\frac 12} Exp \left\{ -\frac{%
m_j\omega _j}{\hbar } \left[ \frac 12 \coth(\beta \hbar \omega_j)
(x_{j0}^2+x_{j0}^{\prime 2} ) -\frac{x_{j0}x_{j0}^{\prime }}{\sinh (\beta
\hbar \omega _j)} \right] \right\} .
\end{equation}
\end{center}

Hence, using eq.(\ref{eqthetaj}) , followed by a straight forward
calculation, we obtain (obviously the arbitrary phase of $\theta _{\xi _j}$
would not influence the result):

\begin{equation}
\rho _j(\xi _j;t)\,=\frac 1{\sqrt{2\pi }\sigma_j}e^{-\frac{\xi _j^2}{%
2\sigma_j^2}},
\end{equation}
which is a gaussian distribution with width:

\begin{equation}  \label{brownwidthksi}
\sigma_j^2=\frac \hbar {2m_j\omega _j}(|b_{j1}(t)|^2+\omega
_j^2|b_{j2}(t)|^2)\,\coth \left( \frac{{\ \hbar }\omega _j}{2kT}\right) .
\end{equation}

The density matrix traced over the variables $\xi _j$ is given by: 
\begin{equation}  \label{density}
\rho (q_1,q_2;t)= \underbrace{\int \int \int...\int }_N \psi
^{*}(q_1-\sum_j\xi _j,t)\psi (q_2-\sum_j\xi _j,t)\prod_{j=1}^N \frac 1{\sqrt{%
2\pi }\sigma_j}e^{-\frac{\xi \,_j^2}{2\sigma_j^2}}d\xi _j.
\end{equation}
Notice that this is not a reduced density matrix in the ordinary sense,
because the meaning of its off diagonal elements is not clear, mainly due to
the fact that the variable $\xi _j(t)$'s are not pure environment variables
unless t=0. However, the diagonal matrix elements $\rho (q,q;t)$ have clear
physical meaning: they are the probability density $\rho (q;t)$. Hence we
shall only address the diagonal elements in the following, even though the
results for the off-diagonal elements are as simple as the diagonal elements.

It is easy to derive the following formula for an arbitrary function $f(\xi )
$: 
\begin{equation}
\underbrace{\int \int \int ...\int }_Nf(\sum_j\xi _j)\prod_{j=1}^N\left(
\frac 1{\sqrt{2\pi }\sigma _j}e^{-\frac{\xi _j^2}{2\sigma _j^2}}d\xi
_j\right) =\frac 1{\sqrt{2\pi }\sigma _\xi }\int f(\xi )e^{-\frac{\xi ^2}{%
2\sigma _\xi ^2}}d\xi \;,
\end{equation}
with 
\begin{equation}
\sigma _\xi ^2={\sum_{j=1}^N}\sigma _j^2.  \label{brownwidth}
\end{equation}

Therefore we have a very simple expression for the probability density:

\begin{equation}
\rho (q,t)=\frac 1{\sqrt{2\pi }\sigma _\xi }\int |\psi (q-\xi ,t)|^2e^{-%
\frac{\xi ^2}{2{\sigma _\xi }^2}}d\xi .  \label{density1}
\end{equation}
where $\sigma _\xi $ is the Brownian motion width, calculated by
substituting eq.(\ref{brownwidthksi}) into eq.(~\ref{brownwidth}): 
\begin{equation}
\sigma _\xi ^2(t)=\int\limits_0^\infty \frac \hbar {2m_j\omega
_j}(|b_{j1}(t)|^2+\omega _j^2|b_{j2}(t)|^2)\,\coth \left( \frac{{\ \hbar }%
\omega _j}{2kT}\right) \rho (\omega _j)d\omega _j,  \label{brownwidth1}
\end{equation}
and by replacing the sum over the bath oscillator number j with the spectral
density $\rho (\omega _j)$ of the bath oscillators as given by eq.(6) of [1]:

\begin{equation}
\rho (\omega _j)=\frac{2\eta M}\pi \frac{m_j\omega _j^2}{c_j^2}.
\end{equation}

This width is zero initially, then approaches its final equilibrium value in
a time interval of the order of 1/$\eta $. In low temperature limit, the
equilibrium width is simplified to: 
\begin{equation}
\sigma_\xi^2 (t=\infty )=\frac \hbar {2\pi M\omega }\left[ \frac \pi
2+\arctan \left( \frac{\omega _0^2}{\eta \omega }\right) \right] \,\,.
\end{equation}
If $\eta \ll \omega _0$, this width happens to become the same as the width
of the ground state of the system $\sigma _0^2=\hbar /(2M\omega _0) $.

There are some subtleties to be addressed in future communications in the
calculation of $\sigma _\xi (t)$ for finite t. They are associated with a
logarithmic divergence of the integration over $\omega _j$ in eq.(\ref
{brownwidth1}). The divergence is removed by introducing a cut-off frequency 
$\omega _j^{(cut)}$ in eq.(\ref{brownwidth1}). In the following, for
simplicity, we consider only the low temperature limit. To get an idea of
the behavior of $\sigma _\xi $, we plot $\sigma _\xi ^2/\sigma _0^2$ in
figure 1, for $\omega _j^{(cut)}=100 \omega$, and $\eta =0.1\omega _0$. As a
rough estimate, we have 
\begin{equation}  \label{brownestimate}
\sigma_\xi^2 \approx \sigma_0^2 (1-e^{-\eta t}).
\end{equation}
This function is not very sensitive to the specific value of $\omega
_j^{(cut)}/\omega _0$.

3. The wavefunction evolution. We assume that initially the wavefunction is
a gaussian wave packet with minimum uncertainty and displaced from the
center of the potential well by z:

\begin{equation}
\psi _0(q)\equiv \psi (q;t=0)=(2\pi \sigma _0^2)^{-1/4}e^{\displaystyle -%
\frac{(q-z)^2}{4\sigma _0^2}}.  \label{iniwave1}
\end{equation}
Using the Green's function (derived as eq.(21) in [1] ):

\begin{equation}
G(Q_1,Q_0;t,0)=\left( \frac{M\omega e^{\frac 12\eta t}}{2\pi i\hbar \sin
\omega t}\right) ^{\frac 12}exp\left[ \frac{iM}{2\hbar a_2}\left(
a_1Q_0^2+\dot a_2e^{\eta t}Q_1^2-2Q_0Q_1\right) \right] .
\end{equation}
we have: 
\begin{equation}
\psi (q,t)=\int \psi (q_0,0)G(q,q_0;t,0)dq_0  \label{wavefunction1}
\end{equation}
\begin{equation}
=\frac{(2\pi \sigma _0^2)^{-1/4}}{\sqrt{a_1+i\omega _0a_2}}exp\left\{ -\frac
1{4\sigma _0^2(a_1+i\omega _0a_2)}\left[ ({\dot a_2}+i\omega _0a_2)e^{\eta
t}q^2-2qz+a_1z^2\right] \right\} ,  \label{wavefunction}
\end{equation}
where $a_1$ and $a_2$ are the well known coefficients of the initial values $%
q_0$ and $\dot q_0$ for the solution of the classical damping equation $%
\ddot q+\eta \dot q+\omega _0^2q=0\;\,$[1]. If $\eta \ll \omega _0$, we
rewrite $a_1$ and $a_2$ (eq.(12) of [1]) approximately as: 
\begin{equation}
a_1=e^{-\frac \eta 2t}(\cos \omega t+\frac \eta {2\omega }\sin \omega
t)\approx e^{-\frac \eta 2t}\cos \omega t
\end{equation}
\begin{equation}
\omega _0a_2=e^{-\frac \eta 2t}\frac{\omega _0}\omega \sin \omega t\approx
e^{-\frac \eta 2t}\sin \omega t,\;\;\dot a_2(t)\approx a_1(t).
\label{a2approx}
\end{equation}
With this approximation, the wave function takes a simple and familiar form: 
\begin{equation}
\psi (q,t)\approx C(t)e^{\displaystyle -\frac{[q-q_c(t)]^2}{4\sigma
_0^2e^{-\eta t}}}e^{ik_c(t)q}.
\end{equation}
This is a gaussian wave packet centered at the classical damped orbit: $%
q_c(t)=e^{-\frac \eta 2t}z\cos \omega t$, with momentum of 
\begin{equation}
\hbar k_c(t)=\hbar \frac z{2\sigma _0^2}e^{\frac \eta 2t}\sin \omega t\simeq
Me^{\eta t}\left[ \frac d{dt}q_c(t)\right] .  \label{wavenumber}
\end{equation}

This corresponds to the wave number of a particle of mass $Me^{\eta t}$. The
coefficient $C(t)$ is simply a normalization constant (independent of q).
Notice that even though the velocity of the classical motion $q_c(t)$
deceases as $e^{-\eta t/2}$, the wavelength $2\pi /k_c(t)$ still decays as $%
e^{-\eta t/2}$ because the effective mass increases exponentially. Notice
also that the width of the wave packet $\sigma _0e^{-\eta t/2}$ deceases
exponentially, so that the probability density distribution evolves into a $%
\delta $ function if Brownian motion can be neglected. Obviously, as time
approaches infinity, the system approaches a classical limit. To provide a
more clear picture of the evolution process, in Figure 2, we plot the real
part of the wavefunction at 4 different times, assuming that the damping
rate is $\eta =0.1\omega _0$, and the initial displacement is $z=5\sigma _0$.

4. The interference fringe wavelength. Now, we are ready to check if the
wavelength decay is observable. If initially the wavefunction is a
superposition of two such wave packets as eq.(\ref{iniwave1}), one with z=0,
the other with z$\neq $0, the above results can be used to calculate
straightforwardly the interference term (which has been studied in detail by
CL using path integral technique[2,4]). The result will show how the
Brownian motion erases the interference.

The initial wavefunction is: 
\begin{equation}  \label{iniwave}
\psi _0(q)=\psi _{01}(q)+\psi _{02}(q)=C(e^{\displaystyle -\frac{q^2}{%
4\sigma _0^2}}+e^{\displaystyle -\frac{(q-z)^2}{4\sigma _0^2}}),
\end{equation}
where $C$ is a normalization constant.

Using eq.(\ref{iniwave}), eq.(\ref{wavefunction}), and eq.(\ref{density1}),
we find: 
\begin{equation}
\rho (q,t)={C}^2\frac{\sigma _0}{\sigma _\theta }\left[ e^{-\frac{q^2}{%
4\sigma _\theta ^2}}+e^{-\frac{(q-a_1z)^2}{4\sigma _\theta ^2}}+e^{-\frac{%
q^2+(q-a_1z)^2}{4\sigma _\theta ^2}}2\cos \left( \frac{q^2-(q-a_1z)^2}{%
4\sigma _\theta ^2}\frac{\omega _0a_2}{a_1}\right) e^{-\frac{z^2}{8\sigma
_0^2}\frac{\sigma _\xi ^2}{\sigma _\theta ^2}}\right]   \label{inter}
\end{equation}
where 
\begin{equation}
{\sigma _\theta ^2}\equiv \sigma _0^2({a}_1^2+{\omega _0^2}a_2^2)+{\sigma
_\xi ^2}.  \label{fullwidth}
\end{equation}
The last term in the parenthesis is the interference term, which agrees with
the equation (2.9) of [4]. The notation is slightly different. For example,
a part of the exponent of the last factor has been identified as the ratio
of the Brownian motion width $\sigma _\xi ^2$ over the full width $\sigma
_\theta ^2$ of the wave packet; $a_1$ is identified as the trajectory of a
particle with initial position 1 but zero velocity, while $a_2$ corresponds
to a particle with initial velocity 1 but position at the origin[1]. In
every period, the maximum interference occurs when $a_1=0,(\omega t\approx
(2n+1)\pi /2)$, as is indicated by the first factor of the interference
term. In the following, we only discuss the time when $a_1(t)=0$.

Since the wavelength decays exponentially in eq.(\ref{wavefunction}), it is
interesting to see whether the interference fringe wavelength also decays
exponentially, and provides an observable phenomenon. We assume $\eta \ll
\omega _0$, use the approximation eq.(\ref{a2approx}), and estimate the
Brownian width by eq.(\ref{brownestimate}). Then the growth of $\sigma _\xi
^2$ and the decay of $a_2^2$ happen to cancel each other, and $\sigma
_\theta ^2\approx \sigma _0^2$ is nearly constant. The eq.(\ref{inter})
becomes: 
\begin{equation}
\rho (q,t)\approx 2{C}^2e^{-\frac{q^2}{4\sigma _0^2}}\left[ 1+\cos \left(
k_{c0}e^{-\frac 12\eta t}q\right) e^{-\frac 12{k_{c0}^2\sigma _\xi ^2}%
}\right] ,  \label{inter2}
\end{equation}
where $k_{c0}\equiv \frac z{2\sigma _0^2}\approx k_c(\omega t=\frac \pi 2)$
(see eq.(\ref{wavenumber})). In order to have enough fringes the initial
wavelength $2\pi /k_{c0}$ should be much smaller than the wave packet width $%
\sigma _0$. Hence the Brownian width increases to the initial wavelength in
a time much shorter than the damping time $1/\eta $ (the time for it to
increase to $\sigma _0$ ). Therefore the eq.(\ref{inter2}) shows that the
fringe wavelength does not decay but increases slightly before the Brownian
motion erases the interference pattern. Thus, in this well studied case, the
exponential decay of wavelength is not experimentally observable!

5. How to observe the wavelength decay. If we carefully examine the
calculation of the probability density eq.(\ref{density1}), we realize that
the slight increase of fringe wavelength is due to the narrow wave packet
width ${\sigma _0^2}$ and the convolution of the wavefunction with the
Brownian motion in eq.(\ref{density1}). To see whether the wavelength decay
is really observable, in the initial wavefunction eq.(\ref{iniwave}) we
replace $\sigma _0$ by a different width ${\sigma }$. Let the ratio $\sigma
_0/\sigma \equiv r$. Now, with a straightforward calculation as before, the
probability density becomes: 
\begin{equation}  \label{inter1}
\rho (q,t)={C}^2\frac \sigma {\sigma _\theta }\left[ e^{-\frac{q^2}{4\sigma
_\theta ^2}}+e^{-\frac{(q-a_1z)^2}{4\sigma _\theta ^2}}+e^{-\frac{%
q^2+(q-a_1z)^2}{4\sigma _\theta ^2}}2\cos \left( \frac{q^2-(q-a_1z)^2}{%
4\sigma _\theta ^2}\frac{\omega _0a_2}{a_1}r^2\right) e^{-\frac{z^2}{8\sigma
^2}\frac{\sigma _\xi ^2}{\sigma _\theta ^2}}\right] ,
\end{equation}
while 
\begin{equation}  \label{sigmatheta}
{\sigma _\theta ^2}\equiv \sigma ^2({a}_1^2+r^4{\omega _0^2}a_2^2)+{\sigma }%
_\xi ^2.
\end{equation}
Again, we use eq.(\ref{brownestimate}) to calculate the fringe wavenumber
(the coefficient of q in the phase of the cosine function in eq.(\ref{inter1}%
)). The fringe wavenumber is then found to grow exponentially until $t\simeq
ln(r^2 -1)/\eta $, before it starts to decrease, due to the Brownian motion
and the finite wave packet width. In the previously studied case, r=1, hence
the wavelength decay is not observable. However, if r is sufficiently large,
it becomes observable, as can be seen as follows. Now, for quite a long
time, $\sigma _\xi ^2$ can be neglected in eq.(\ref{sigmatheta}), while $%
\sigma _\theta ^2$ decays exponentially, the interference pattern can then
be simplified to (when $a_1=0$ and when the approximation eq.(\ref{a2approx}%
) is valid): 
\begin{equation}
\rho (q,t)={C}^2\frac{2\sigma }{\sigma _\theta }e^{-\frac{q^2}{4\sigma
_\theta ^2}}\left[ 1+\cos \left( k_c(t)q\right) e^{-\frac 12{k_c^2}(t){%
\sigma _\xi ^2}}\right] .
\end{equation}

In the mean time, the visibility of the fringes also decays. The time when
it is reduced by a factor of e is determined by $\frac{{k_c}^2(t)\sigma _\xi
^2}2=1$, i.e., when the wavelength is about equal to the Brownian motion
width. Using eq.(\ref{brownestimate}), we find the time is $t\simeq ln(1+8{%
\sigma _0}^2/z^2)/\eta $ . Thus large r and small z tend to give longer time
to observe the wavelength decay before the fringes disappear. The number of
visible fringes in the wave packet is $2k_c(t)\sigma _\theta \simeq z/\sigma 
$ when $t\simeq (2n+1)\pi /2$, so z should be larger than $\sigma $ to have
enough fringes. In figure 3 we plot the probability density at different
times for a case with $z=3\sigma _0$ and r =16. A comparison of fig.3b and
fig.3c clearly shows the wavelength decay and the visibility decay. In
fig.3d, the coherence has been erased almost completely, long before the
distribution width becomes $\sigma _0$.

\begin{center}
{\bf Acknowledgments}
\end{center}

The author thanks Prof. C.N. Yang for many sessions of stimulating
discussions. The author also likes to thank C.P. Sun for interesting
discussions. The work is performed under the auspices of the U.S. Department
of Energy under Contract No. DE-AC02-76CH00016.

\newpage

\begin{center}
{\bf References}
\end{center}

\vspace{0.5cm}

\begin{enumerate}
\item  L.H.Yu, C.P.Sun, Phys. Rev. A {\bf 49}, 592 (1994)

\item   A.O. Caldeira, A.J. Leggett, Ann. Phys. 149, 374 (1983), and Physica
121A, 587 (1983)

\item  E. Kanai, Prog. Theor. Phys. 3, 440 (1948); S. Nakajima, Prog. Theor.
Phys. 20, 948 (1958); R.Zwanzig, J. Chem. Phys. 33, 1338 (1960); I.R.
Senitzky, Phys. Rev. 119, 670 (1960); G.W. Ford, M. Kac, P. Mazur, Jour.
Math. Phys., 6, 504 (1965); P. Ullersma, Physica {\bf 32} (1966); M. D.
Kostin, J. Chem. Phys. 57, 3589 (1972); K. Yasue, Ann. Phys. 114, 479
(1978); R.H. Koch, D.J. Van Harlingen, J. Clarke, Phys. Rev. Lett, 45, 2132
(1980); H. Dekker, Phys. Report, 80, 1 (1981); R. Benguria, M. Kac, Phys.
Rev. Lett, 46, 1 (1981)

\item   A.O. Caldeira, A.J. Leggett, Phys. Rev. A, 31, 1059 (1985)

\item  R.P.Feynman, Statistical Mechanics, W.A.Benjamin, Inc. (1972)

\vspace{0.5cm}
\end{enumerate}

\newpage

\begin{center}
{\bf Figure Captions}
\end{center}

Figure 1. The Brownian motion width as a function of time. The cut-off
frequency $\omega _j^{(cut)}=100 \omega$, $\eta =0.1\omega _0$.

Figure 2. The time evolution of $\psi (q,t)$. $z=5\sigma _0,\eta =0.1\omega
_0$.

Figure 3. The time evolution of the interference pattern. The probability
density is in arbitrary unit. $z=5\sigma _0,\eta =0.1\omega _0,r=16,\omega
_j^{(cut)}=100 \omega$.

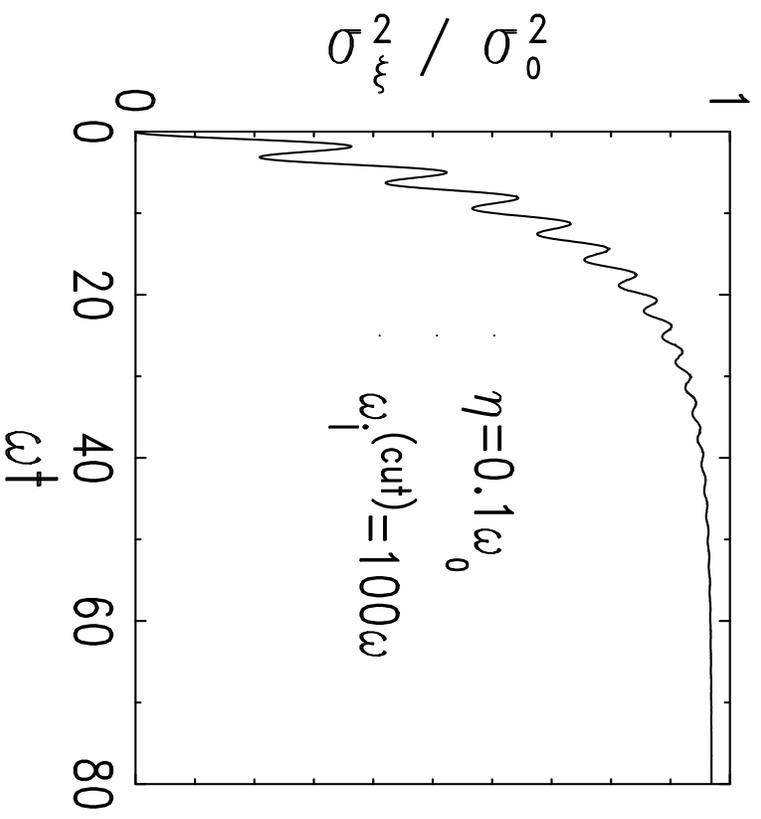
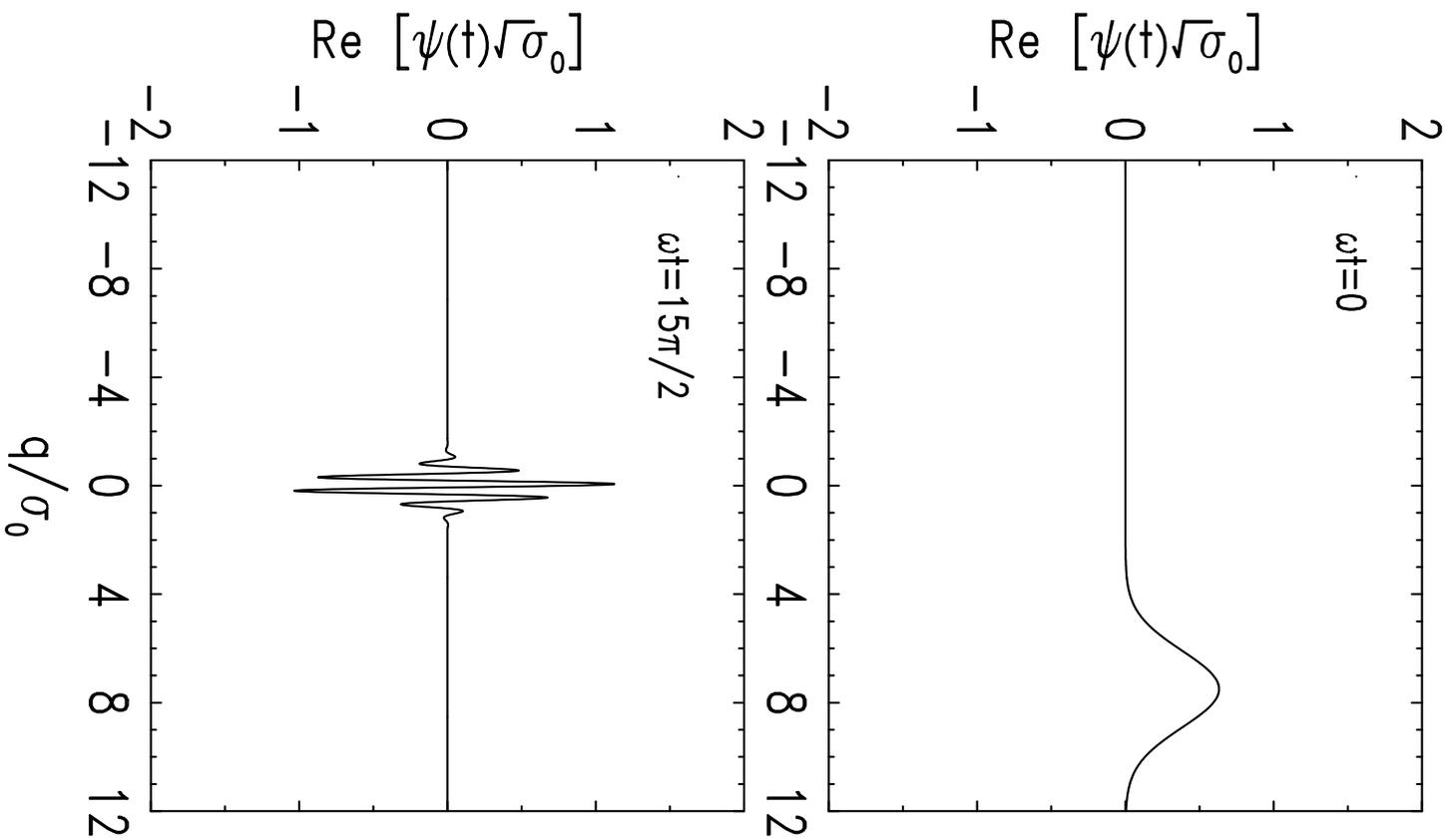
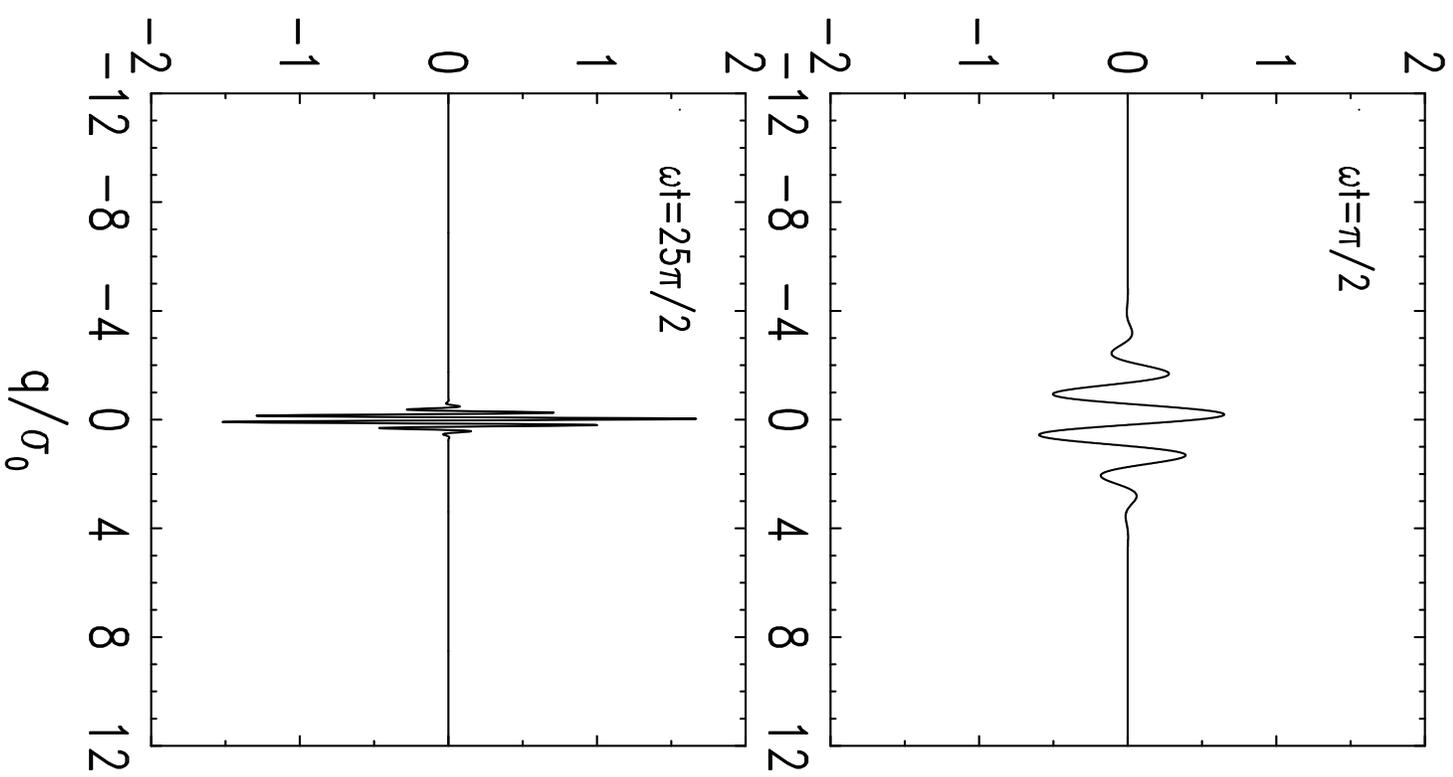
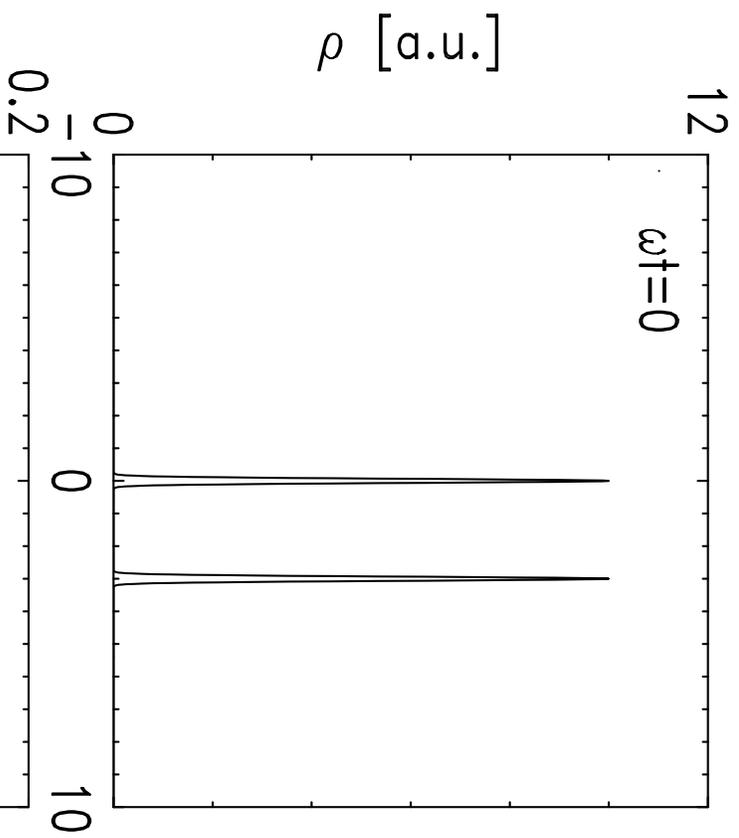
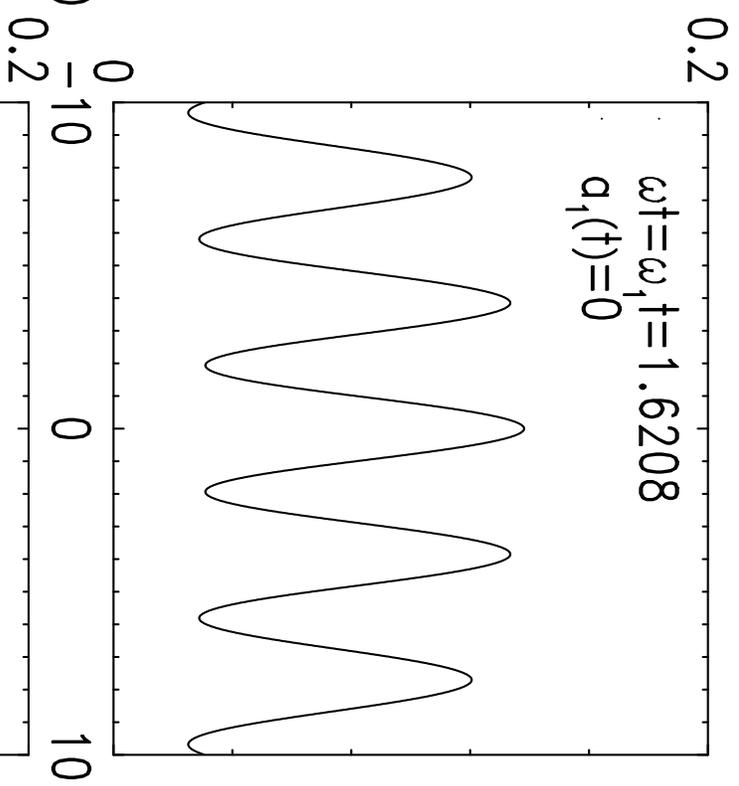
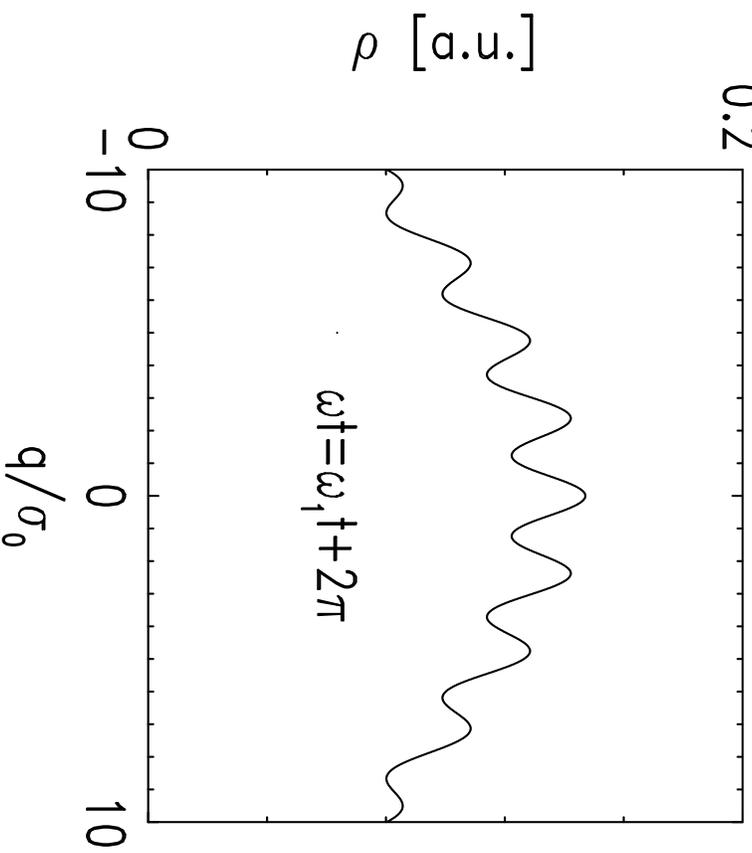
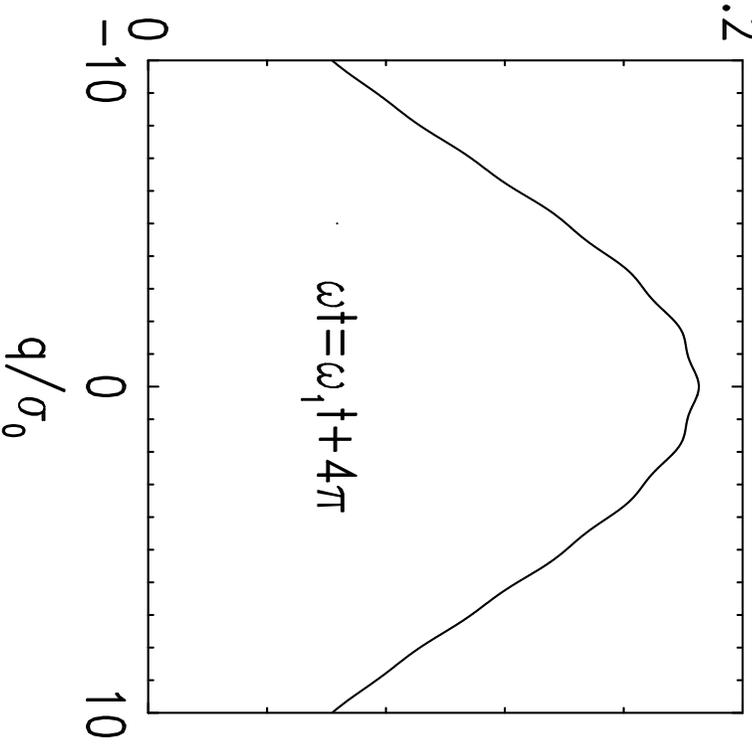
\end{document}